# Gamma Ray Large Area Space Telescope (GLAST) Balloon Flight Data Handling Overview

T. H. Burnett, A. Chekhtman, E. do Couto e Silva, R. Dubois, D. Flath, I. Gable, J. E. Grove,
R. Hartman, T. Kamae, A. Kavelaars, H. Kelly, T. Kotani, M. Kuss, D. Lauben, T. Lindner, N. Lumb,
T. Mizuno, A. Moiseev, M. Ozaki, L. S. Rochester, R. Schaefer, G. Spandre, D. J. Thompson,
T. Usher, and K. Young, *on behalf of the GLAST Large Area Telescope Collaboration*

*Abstract--* **The GLAST Balloon Flight Engineering Model (BFEM) represents one of 16 towers that constitute the Large Area Telescope (LAT), a high-energy (>20 MeV) gamma-ray pair-production telescope being built by an international partnership of astrophysicists and particle physicists for a satellite launch in 2006. The prototype tower consists of a Pb/Si pair-conversion tracker (TKR), a CsI hodoscopic calorimeter (CAL), an anti-coincidence detector (ACD) and an autonomous data acquisition system (DAQ). The self-triggering capabilities and performance of the detector elements have been previously characterized using positron, photon and hadron beams. External target scintillators were placed above the instrument to act as sources of hadronic showers. This paper provides a comprehensive description of the BFEM data-reduction process, from receipt of the flight data from telemetry through event reconstruction and background rejection cuts. The goals of the ground analysis presented here are to verify the functioning of the instrument and to validate the reconstruction software and the background-rejection scheme.**

Manuscript submitted November 5, 2001.

This work is supported in part by Department of Energy contract DE-AC03-76SF00515 and NASA contract NAS598039.

L. S. Rochester, the corresponding author, is with the Stanford Linear Accelerator Center, 2575 Sand Hill Road, MS 78, Menlo Park, CA 94025 USA (telephone: 650-926-2695, e-mail: leon.rochester@slac.stanford.edu) as are E. do Couto e Silva, R. Dubois, T. Kamae, A. Kavelaars, T. Usher and K. Young.

T. H. Burnett is with the University of Washington, Seattle WA 98195 USA.

A. Chekhtman and J. E. Grove are with the Naval Research Laboratory., Washington, DC 20375 USA.

D. Flath, I. Gable and T. Lindner are with the University of Victoria, Victoria, BC V8W 2Y2, Canada.

R. Hartman, H. Kelly, T. Kotani, A. Moiseev, R. Schaefer and D. J. Thompson are with  NASA Goddard Space Flight Center, Greenbelt MD 20771 USA.

M. Kuss and G. Spandre are with the University of Pisa, INFN Pisa, 56010 S. Piero a Grado (PI), Italy.

D. Lauben is with Stanford University, Stanford CA 94305 USA.

N. Lumb is with Université Claude Bernard Lyon-I, IPNL, F-69622 Villeurbanne, France.

T. Mizuno is with  Hiroshima University, Higashi-Hiroshima, 739-8511, Japan.

M. Ozaki is with the Institute of Space and Aeronautical Sciences, Sagamihara, 229-85, Japan.

## I. INTRODUCTION

THE Balloon Flight Engineering Model (BFEM) consists of three major detector components: the tracker (TKR), a stack of silicon strip detectors and thin lead foils; the calorimeter (CAL), an array of CsI(Tl) logs; and the anti-coincidence detector (ACD), an array of plastic scintillators covering the tracker. It represents one of the sixteen towers that constitute the Large Area Telescope (LAT) of the Gamma ray Large Area Space Telescope (GLAST) [1]. The LAT is designed to detect photons with energies above about 20 MeV, and up to several hundred GeV. Cosmic gamma rays pass undetected through the ACD, convert to electron-positron pairs in the tracker and deposit their energies in the calorimeter. The instrument triggers when a track segment traverses any three contiguous silicon layers, which are composed of two silicon planes, each measuring one direction perpendicular to the tower axis.

The raw data of this telescope are not images. Because the LAT is designed for pair-production energies, it is basically a particle detector. Photons are detected as discrete "events" consisting of the tracks (energy deposit) left by ionizing particles in different parts of the instrument. These events need to be classified as incident charged particles or photon-induced particle pairs. The quality of the track reconstruction determines our ability to resolve sources, and our ability to separate charged particles from gamma-ray pair events determines the level of background contaminating our sample. An overview of the BFEM development and flight program, including results on subsystem performance, is published separately [2].

The ground-based analysis starts with the raw data recorded through the telemetry system on the balloon. Error checking is performed and then the data are converted to a form convenient for further processing. The converted data are used as input to a reconstruction package that finds charged tracks in the tracker and clusters of energy deposition in the calorimeter. The tracks are extrapolated to the ACD to determine whether a tile fired along the path of the track.

*Work supported in part by Department of Energy Contract DE-AC03-76SF00515.
Stanford Linear Accelerator Center, Stanford University, Stanford, CA  94309, USA

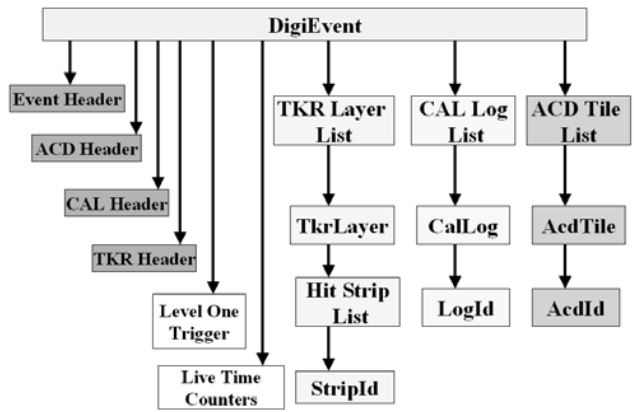

Fig. 1. Logical structure of the raw detector data stored in ROOT.

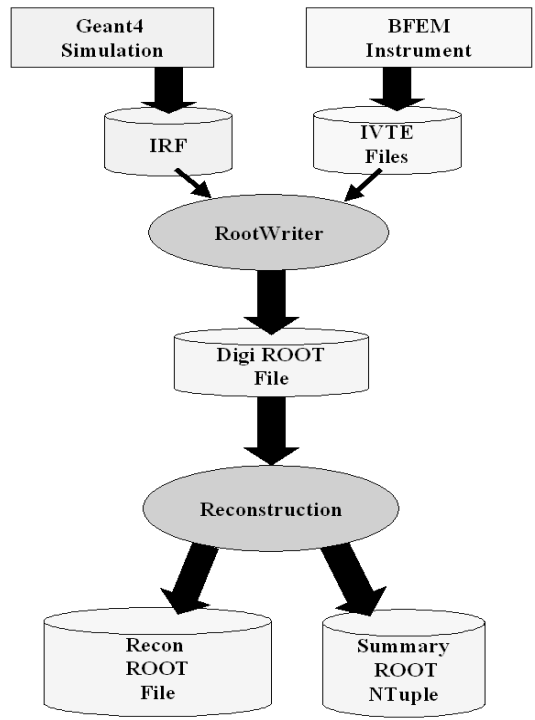

Fig. 2. Data flow from simulations and BFEM instrument through reconstruction.

Potential photons, consisting of two tracks in a "vee" (or possibly a single track, for some high energy photons), are written to an output file, where they are available for further analysis. At this stage, cuts can be made to reject background events.

The data-handling sequence described here is similar to the one that will be used in the final flight instrument. Indeed, one of the purposes of the balloon flight was to allow us to validate and refine the data analysis.

## II. PREPARATION OF INPUT DATA

The balloon flight provided an opportunity to test our data processing and analysis facilities. Much of our existing software, including the Monte Carlo simulation and event reconstruction routines, is written in C++. A file format that adheres to the object-oriented paradigm reduces the complexity of reading and writing event data. An I/O and analysis package called ROOT [3] allows us to store C++ objects inside a file. The ROOT I/O package is designed to create compact files, as well as allow for efficient access to the data. The tree structure of the ROOT files allows a subset of the branches to be manipulated, reducing the amount of I/O required. For example, only a small fraction of the flight data is from gamma rays; a simple C++ script will extract those likely photon events and create a new truncated ROOT file.

Our intention is that the ROOT files containing data, whether from the actual balloon flight or our simulation of it, or from any of the other incarnations of the instrument, including the final one launched into orbit, all have the same internal structure. Hence, I/O and low-level analysis routines can be shared. This will greatly minimize the programming effort, as the same functions will not have to be rewritten for each data source. We currently store detailed Monte Carlo truth, detector digitization, and reconstruction data in ROOT files. For example, Fig. 1 illustrates the logical tree structure for the detector digitization data.

The BFEM generated Integrated Variable-Length Tower Event (IVTE) files, which contain the detector digitization data. These files were checked for integrity and then converted to ROOT format by a program called RootWriter.

RootWriter can convert BFEM IVTE files, as well as instrument response files (IRF) from our GEANT4 [4], [5] Monte Carlo simulation into ROOT files. The various formats are converted into C++ objects that are then stored in a ROOT file within its tree structure. The same event structure is used for both real and simulated data, allowing easy comparison between the two. After processing the digitization data, the reconstruction routines produce a Recon ROOT file and a summary file, also in ROOT format, containing the results of the reconstruction.

Fig. 2 provides a diagram of the data flow from both the Monte Carlo simulation and the BFEM instrument.

Data analysis can then be performed using the ROOT analysis package, which includes graphics capabilities. A ROOT-based event display was created and is used to scan the events. For those interested in using Interactive Data Language (IDL) [6], which is widely used in the astrophysics community for data analysis and visualization, a program called Root2IDL converts ROOT objects into IDL structures.

## III. EVENT RECONSTRUCTION

The event reconstruction takes the digitizations from the detector elements, converts them to physics units (e.g. energies in MeV, distances in mm), performs pattern recognition and fitting to find tracks and then photons in the tracker, finds energy clusters in the calorimeter and characterizes their energies and directions. Tracks that extrapolate to a fired ACD tile can be identified. Many of the techniques discussed below were developed during tests of a similar instrument in particle beams at the Stanford Linear Accelerator Center. [7], [8].

### A. Tracker

The tracker consists of a tower of silicon strip detectors, arranged in pairs, with each element of the pair providing a separate measurement in one direction (X or Y) perpendicular to the tower axis. The tracker reconstruction is initially done in the separate Z-X and Z-Y projections. The projections are associated with each other whenever possible by matching tracks with respect to length and starting positions.

In the absence of interactions, particle trajectories through our detector would be straight lines. However, the converter foils, needed to produce the interactions, as well as the rest of the material in the detector, cause the particles to undergo multiple Coulomb scattering (MS) as they traverse the tracker. This complicates both the pattern recognition (finding the particles) and track fitting (determining the particle trajectories), particularly for low-energy electrons and positrons.

Thus, our pattern recognition must be sensitive to particles whose trajectories depart significantly from straight lines. The presence of multiple scattering also has implications for the fitting procedure. Without MS, deviations from a straight line are due solely to measurement errors, which occur independently at each measurement plane, and are distributed about the true straight track. In the presence of MS, there are real random deviations from a straight line, and these deviations are correlated from one plane to the next. For example, if an individual particle scatters to the right at one plane, it is more likely to end up to the right of the original undeviated path than to the left.

These correlations can be quantified in a *covariance matrix* of the measurements, which is calculated from the momentum of the particle and the amount of scattering material between the layers. The dimension of this matrix is the number of measurements. Solving for the track parameters in terms of the measurements involves inverting this matrix. In the case of no MS, the matrix is diagonal and the inversion is trivial; MS introduces off-diagonal elements, which complicates the inversion.

Another technique, the Kalman Filter (KF) [9], can be useful in both stages of particle reconstruction. This starts with an estimate of the initial position, direction and energy of the particle. In our case, the energy of the particle is estimated from the response of the calorimeter (see below), and the starting point and direction come from looking for three successive hits that line up within some limits. From this starting point, the track is extrapolated in a straight line to the next layer. Using the estimate of the energy and the amount of material traversed, we can decide whether the hit in the next layer is within a distance from the extrapolated track allowed by the expected multiple scattering and the uncertainty of the initial estimate. If so, the hit is added to the track, and the position and direction of the track at this plane are modified, incorporating the information from the newly added hit. The modified track is now extrapolated to the next plane, and the process continues until there are no more planes with hits. All the correlations between layers have been properly taken into account, but at each step, only MS between two successive planes need be considered, and the covariance matrix required is that of the parameters, which in this simple case is of dimension two, rather than the much larger one described earlier.

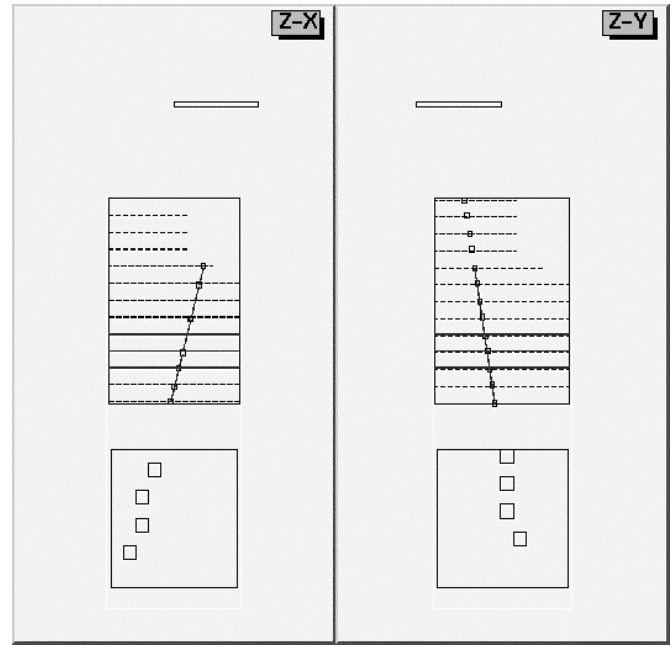

Fig. 3. Fit to a cosmic ray track (solid line) in the tracker. Reconstructed energy centroids in the calorimeter (boxes) line up with the track. The track extrapolates to the fired ACD tile.

The track parameters at the last hit have now been calculated using the information from all the preceding hits. However, we usually want to know the parameters at the first hit, close to the point where the photon converts to an electron-positron pair, to get the best estimate of the initial direction of the photon. To do this, *smoothing* is applied, that is the KF is "run backwards" from the last plane to the first, using the appropriate matrices.

After smoothing, the track parameters, and their errors, have been calculated at each of the measurement planes, and in particular, the first plane.

Fig. 3 shows the result of the fitting algorithm applied to a cosmic ray track.

Stanford Linear Accelerator Center, Stanford University, Stanford, CA 94309, USA

## B. Calorimeter

A high-energy photon traversing material loses its energy by an initial pair-production process ($\gamma \rightarrow e^+ e^-$) followed by subsequent *bremsstrahlung* ($e \rightarrow e\gamma$) and pair production, resulting in an electromagnetic cascade, or shower. The scale length for this shower is the *radiation length*, the mean distance over which a high-energy electron loses all but 1/e of its initial energy.

The calorimeter provides information about the total energy of the shower, as well as the position and direction and shape of the shower, or of the penetrating nucleus or muon. It consists of eight layers of ten CsI(Tl) crystals ("logs") in a hodoscopic arrangement, that is, alternatively oriented in X and Y directions, to provide an image of the electromagnetic shower. It is designed to measure photon energies from 20 MeV to 300 GeV and beyond.

To comfortably contain photons with energies in the 100-GeV range requires a calorimeter at least twenty radiation lengths thick. However, weight constraints forced our calorimeter to be only ten radiation lengths in thickness, and thus it cannot provide good shower containment for these high-energy photons, even though they are very precious for several astrophysics topics. Indeed, the mean fraction of the shower contained at 300 GeV is about 30% for photons at normal incidence. In this case, the energy observed becomes very different from the incident energy, the shower development fluctuations become larger, and the resolution decreases quickly.

Two solutions have been pursued so far to correct for the shower leakage. The first is to fit a mean shower profile to the observed longitudinal profile. The profile-fitting method proves to be an efficient way to correct for shower leakage, especially at low-incidence angles when the shower maximum is not contained. After the correction is applied, the resolution (as determined from our simulation) is 18% for on-axis 1 TeV photons. This is an improvement by a factor of two over the result of correcting the energy with a response function based on path length and energy alone.

The second method uses the correlation between the escaping energy and the energy deposited in the last layer of the calorimeter. The last layer carries the most important information concerning the leaking energy: the total number of particles escaping through the back should be nearly proportional to the energy deposited in the last layer. The

Both the shower-profiling and leakage-correction methods significantly improve the resolution. The correlation method is more robust, since it does not rely on fitting individual showers, but its validity is limited to relatively well-contained showers, making it difficult to use at more than 70 GeV for low-incidence-angle events. There is still some room for improvement in energy reconstruction, especially by correcting for losses in the passive material between the different calorimeter modules and out the sides.

Because of the limited duration of the balloon flight, and the steeply falling energy spectrum of the gamma rays, we expected to detect few if any high-energy photons in the BFEM data. However, the shower-leakage issues discussed above start to become measurable at energies of a few GeV, so they still enter into any detailed analysis of our data.

## IV. EVENT SELECTION

Background rejection performs the function of particle identification, determining whether the incoming particle was a photon. With a charged-particle flux in the upper atmosphere two orders of magnitude larger than the photon flux, and even higher in space, shower fluctuations in background interactions can mimic photon showers in non-negligible numbers. Cuts are applied to the events to suppress the background.

We have implemented a set of simple and intuitive cuts, based partly on previous experience with EGRET [10]. These will not necessarily result in the highest possible efficiency to find photons, but will isolate a set of clean conversion events, and serve to demonstrate that photons can be found, and that background particles can be eliminated.

First, all events are reconstructed as described in section III. We then consider only events in which none of the ACD tiles fired. This cut could be applied before any reconstruction, but reconstructing all the events is useful if we wish to compare the data with simulations. The cut eliminates 90% of triggered events. Most of the rejected events consist of charged particles, but a few legitimate photons will also be rejected if, for example, one of the particles in the shower exits the detector through the sides or top, and fires an ACD tile.

Next, the reconstructed tracks are tested for track quality, formed from a combination of goodness-of-fit, length of track and number of gaps on the track. Also, an energy-dependent cut removes events with tracks that undergo an excessive amount of scattering.

Finally, we require that there be a downward-going vee in both views, and that both tracks in the vee extrapolate to the calorimeter. As noted earlier, this will introduce some inefficiency for high-energy photons, and for highly asymmetric electron-positron pairs. Vees with opening angles that are too large (>60°) are rejected. Such vees generally come from photons with energies below our range of interest.

In the actual balloon flight data, about 0.3% of the triggered events survive all these cuts. A visual inspection verifies that these events appear to be clean photon conversion candidates.

We are developing a number of additional cuts involving extra particles in the event and extra hits not associated with tracks. In addition, we are starting to look at the spatial distribution of energy in the calorimeter, as a way of distinguishing electromagnetic from hadronic showers, and from the showers of photons traveling upward through the instrument.

Fig. 4 shows a photon that converts in the tracker and deposits energy in the calorimeter.

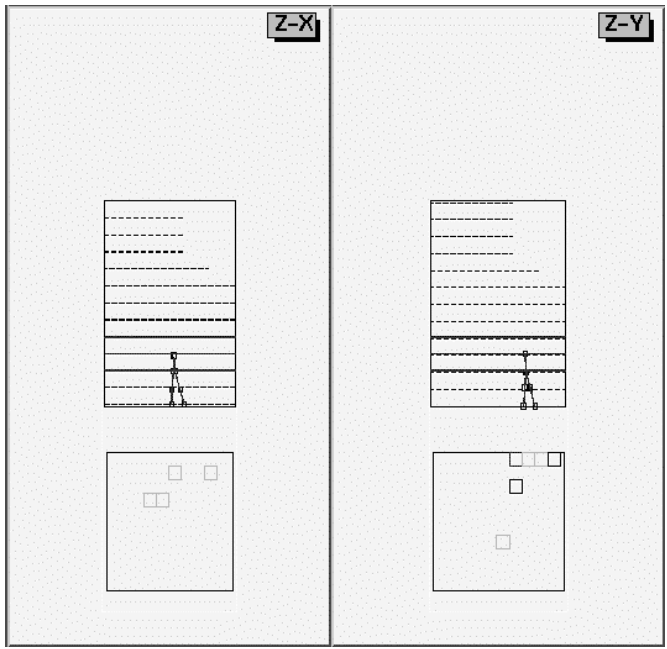

Fig. 4. A reconstructed photon. Note the vee in the tracker, the energy deposited in the calorimeter, and the absence of any signals in the ACD.

## V. CONCLUSION

The data-handling sequence used in processing the data from the BFEM allowed us to verify that the instrument functioned correctly, and that the structures put in place allow us to analyze the data and find photon candidates. This sequence will be used as the basis for the analysis of the data from the full flight instrument.